\documentclass[a4paper,10pt,twoside]{cpc-hepnp}
\usepackage{upgreek,fancyhdr}
\usepackage{multicol}
\usepackage{graphicx}
\usepackage{booktabs}
\usepackage{amssymb,bm,mathrsfs,bbm,amscd}
\usepackage[tbtags]{amsmath}
\usepackage{lastpage}

\begin{document}
%\begin{CJK*}{GB}{gbsn}
%\begin{CJK*}{GBK}{song}

\fancyhead[c]{\small Chinese Physics C~~~Vol. xx, No. x (201x) xxxxxx}
\fancyfoot[C]{\small 010201-\thepage}

\footnotetext[0]{Received 31 June 2015}

\title{The self-consistent analysis for sub-barrier fusion enhancement effect in Ca + Ca and Ni + Ni\thanks{Supported by the National Key Basic Research Development Program of China under Grant No.~2013CB834404, and the National Natural Science Foundation of China under Grants No.~11475263, No.~11375268, No.~U1432246 and No.~U1432127. }}

\author{%
      Nan-Ru Ma (ÂíÄÏÈã)
\quad Hui-Ming Jia (¼Ö»áÃ÷)$^{1)}$\email{jiahm@ciae.ac.cn, corresponding author}%
\quad Cheng-Jian Lin (Áֳмü)$^{2)}$\email{cjlin@ciae.ac.cn, corresponding author}%
\quad Lei Yang (ÑîÀÚ)\\
\quad Xin-Xing Xu (ÐìÐÂÐÇ)
\quad Li-Jie Sun (ËïÁ¢½Ü)
\quad Feng yang (Ñî·å)
\quad Zhen-Dong Wu (ÎâÕñ¶«)\\
\quad Huan-Qiao Zhang (ÕÅ»ÀÇÇ)
\quad Zu-Hua Liu (Áõ×滪)
\quad Dong-Xi Wang (Íõ¶«çô)
}
\maketitle

\address{%
China Institute of Atomic Energy, Beijing 102413, China\\
}

\begin{abstract}
The fusion dynamic mechanism of heavy-ions at energies near the Coulomb barrier is complicated and still not very clear up to now.
Accordingly, a self-consistent method based on the CCFULL calculations has been developed and applied for an ingoing study of the effect of the positive $Q$-value neutron transfer (PQNT) channels in this work.
The typical experimental fusion data of Ca + Ca and Ni + Ni is analyzed within the unified calculation scheme.
The PQNT effect in near-barrier fusion is further confirmed based on the self-consistent analysis and extracted quantitatively.
\end{abstract}



\footnotetext[0]{\hspace*{-3mm}\raisebox{0.3ex}{$\scriptstyle\copyright$}2013
Chinese Physical Society and the Institute of High Energy Physics
of the Chinese Academy of Sciences and the Institute
of Modern Physics of the Chinese Academy of Sciences and IOP Publishing Ltd}%

\begin{multicols}{2}

\section{Introduction}

Low-energy nuclear fusion with heavy-ions offers a fine explorer for investigating the dynamics of the many-body systems.
This process also relates to the current synthesis mechanism of the superheavy elements (SHE)~\cite{Bao15} and has important implication for nucleosynthesis in astrophysics~\cite{Gasques07}.
Therefore, this reaction process has attracted plenty of research in the last decades~\cite{Canto15, Back14, Dasgupta98, Balantekin98, Hagino12, Beckerman80, Montagnoli12, Trotta01, Jiang10, Zhang10, Jia12, Jia14, Broglia83-C, Umar06, Hagino15, Wang14, WB-SCG16, Lin03}.

For this process, a feature is the involvement of the couplings between the relative motion of the colliding nuclei and the intrinsic degrees of freedom such as low-lying collective vibrations of the target and projectile as well as the nucleon transfers ~\cite{Dasgupta98}.
Up to now, the coupled-channel (CC) model has described successfully the fusion enhancement phenomenon correlated with the collective excitations~\cite{Balantekin98,Hagino12}.
While the coupling effect of the positive $Q$-value neutron transfer (PQNT) channels involved in this process, which was first discovered in the experimental study for the near-abrrier fusion excitation functions of $^{58,64}$Ni + $^{58,64}$Ni~\cite{Beckerman80}, is complicated.
Up to now, the relevant experimental conclusions for this effect are still inconsistent and the theoretical descriptions are still immature.
On the other hand, some proposed reduction methods~\cite{Canto15} also cannot successfully give a consistent systematic behavior for the relevant coupling effect.

For solving this problem, one is confronted with the inconsistent experimental data for the same systems measured by different groups.
At energies below the Coulomb barrier, the discrepancy of the measured fusion cross sections even reaches several times in some cases.
Theoretically, problems also exists in the current analysis by the different methods.
For example, although the CC model give a good account for the collective coupling effect, the CC method presents limitations such as a need for external parameters to describe the nucleus-nucleus potential and the couplings~\cite{Simenel13}.

Taking these into account, very recently, a self-consistent RE method was proposed~\cite{Jia16} for trying to disentangle the PQNT coupling effect from the inelastic coupling effect in fusion based on the CCFULL calculation, with the coupling schemes for the collective excitations determined from the experimental fusion data of the reference systems.
Where RE = $\sigma_{\mathrm{Exp}}$/$\sigma_{\mathrm{CC}}$, $\sigma_{\mathrm{Exp}}$ is the experimental fusion cross section and $\sigma_{\mathrm{CC}}$ is the CC calculation result considering the major inelastic couplings.
The proposed RE method reduces overwhelmingly the uncertainty caused by the inelastic coupling effect in the CC calculation and can obtain a reliable quantitative PQNT effect.

\section{Analysis procedure}

In this work, for further studying the relevant problems in fusion reaction, typical near-spherical examples, $^{40,48}$Ca + $^{40,48}$Ca and $^{58,64}$Ni + $^{58,64}$Ni measured by the same groups respectively, optimally suited for this purpose were selected.
The corresponding $Q_{\mathrm{gg}}$-values for the multi-neutron transfer channels from ground state to ground state are shown in Table~\ref{Q}.

The microscopic calculation for the fusion cross section is performed by using the code CCFULL~\cite{Hagino99}.
For the systems studied in the present work, the deformation parameters of the collective excitation states are obtained from the $B$($E$$\lambda$) transitions in NNDC.
Only one-phonon states of the reactants are considered in the calculations.
The relevant information is shown in Table~\ref{Ex}, where $\lambda^{\pi}$ is spin and parity, $E_{\lambda}$ is excitation energy and $\beta_{\lambda}$ is deformation parameter.
All the mutual excitations are considered in the calculations.
The effect of higher excitation states on sub-barrier fusion is expected to be small due to the adiabatic nature of the fusion process~\cite{Hagino97} and are ignored here.
The coupling radius parameter $r_{\mathrm{0c}}$ = 1.10 fm.
The adopted Aky\"{u}z-Winther (AW) proximity potential~\cite{AW} was parameterized into the Woods-Saxon (WS) form with three parameters $V_{\mathrm{0}}$, $r_{\mathrm{0}}$, and $a$.
The parameters for the WS potential and the uncoupled barrier for the studied systems are given in Table~\ref{Par}.
The experimental data were renormalized according to the calculation results at energies above the Coulomb barriers.
All the values of $\beta_{\lambda}$ for the nuclei obtained by fitting the experimental fusion data of the reference systems are in agreement with those obtained from the reduced transition probability $B(E\lambda)$, except for $^{40}$Ca, which was deduced from Ref.~\cite{Timmers98}.

\begin{center}
\tabcaption{\label{Q} The $Q_{\mathrm{gg}}$-values for the multi-neutron transfer channels. The unit is MeV.}
\footnotesize
\begin{tabular*}{70mm}{crrrr}
\toprule System & $Q_{\mathrm{+1n}}$ & $Q_{\mathrm{+2n}}$ & $Q_{\mathrm{+3n}}$ & $Q_{\mathrm{+4n}}$ \\
\hline
$^{40}$Ca + $^{40}$Ca  &  -7.28 & -9.09 &  -18.12  & -21.78 \\
$^{40}$Ca + $^{48}$Ca  &  -1.58 & 2.62  &  0.16    & 3.88   \\
$^{48}$Ca + $^{48}$Ca  &  -4.80 & -5.72 &  -11.76  & -14.45 \\
$^{58}$Ni + $^{58}$Ni  &  -3.22 & -2.08 &  -10.90  & -14.50 \\
$^{58}$Ni + $^{64}$Ni  &  -0.66 & 3.89  &  1.11    & 3.89   \\
$^{64}$Ni + $^{64}$Ni  &  -3.56 & -1.45 &  -6.23   & -6.26  \\
\bottomrule
\end{tabular*}
\vspace{0mm}
\end{center}
\vspace{0mm}

\begin{center}
\tabcaption{\label{Ex} The parameters used for the considered low-lying collective excitation states in the CC calculations.}
\footnotesize
\begin{tabular*}{50mm}{cccc}
\toprule Nucleus & $\lambda^{\pi}$ & $E_{\lambda}$/MeV & $\beta_{\lambda}$\\
\hline
$^{40}$Ca  &  $3^{-}$ & 3.737 &  0.270  \\
           &  $2^{+}$ & 3.904 &  0.119  \\
$^{48}$Ca  &  $2^{+}$ & 3.832 &  0.104  \\
           &  $3^{-}$ & 4.507 &  0.175  \\
$^{58}$Ni  &  $2^{+}$ & 1.454 &  0.154  \\
           &  $3^{-}$ & 4.475 &  0.135  \\
$^{64}$Ni  &  $2^{+}$ & 1.346 &  0.158  \\
           &  $3^{-}$ & 3.560 &  0.216  \\
\bottomrule
\end{tabular*}
\vspace{0mm}
\end{center}
\vspace{0mm}

\end{multicols}
\ruleup
\begin{center}
\includegraphics[width=5.5in]{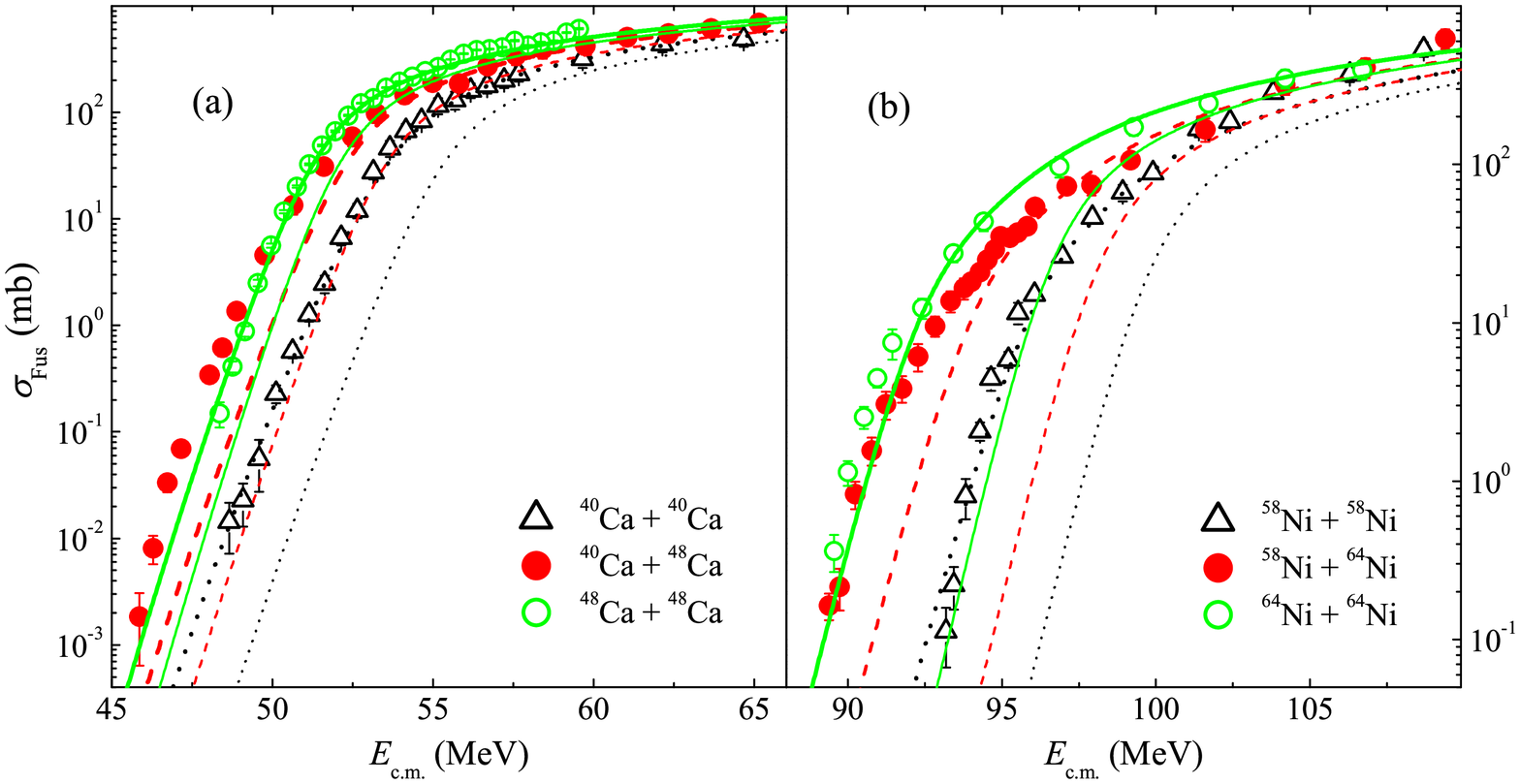}
\figcaption{\label{CS} (Color online)
The CC calculations for the fusion excitation functions of $^{40,48}$Ca + $^{40,48}$Ca (a) and $^{58,64}$Ni + $^{58,64}$Ni (b).
The original experimental fusion data are taken from Refs.~\cite{Beckerman80,Montagnoli12,Trotta01,Jiang10}.}
\end{center}
\ruledown

\begin{multicols}{2}

\section{Results and discussion}

The calculation results for $^{40,48}$Ca + $^{40,48}$Ca with single-channel (SC) and coupled-channels by using the adopted coupling schemes are given in Fig.~\ref{CS}(a).
It can be seen that the experimental fusion cross sections for all the three systems show strong enhancement compared with the SC calculation results (thin lines) at near- and below-barrier energies.
Among them, both $^{40}$Ca + $^{40}$Ca and $^{48}$Ca + $^{48}$Ca have no PQNT channels and were used as a benchmark for the suitable inelastic coupling effects in the CC calculations.
The CC calculations results (thick lines) reproduce qualitatively the experimental fusion excitation functions of the two symmetrical $^{40}$Ca + $^{40}$Ca (dotted line) and $^{48}$Ca + $^{48}$Ca (solid line) systems.
While the CC calculation result, with the coupling schemes extracted from $^{40}$Ca + $^{40}$Ca and $^{48}$Ca + $^{48}$Ca, underestimates much the experimental fusion cross sections of the asymmetrical $^{40}$Ca + $^{48}$Ca system (dashed line) at near- and below-barrier energies.
This is a strong indication of the sub-barrier fusion enhancement correlated with PQNT for $^{40}$Ca + $^{48}$Ca.
The fusion excitation functions of $^{58,64}$Ni + $^{58,64}$Ni in Fig.~\ref{CS}(b) also show a similar behavior.

\end{multicols}
\begin{center}
\tabcaption{\label{Par} The parameters for the WS potential and the uncoupled barrier for the studied systems.}
\footnotesize
\begin{tabular*}{106mm}{ccccccc}
\toprule
System & $V_{\mathrm{0}}$/MeV & $r_{\mathrm{0}}$/fm & $a$/fm & $V_{\mathrm{B}}$/MeV & $R_{\mathrm{B}}$/MeV & $R_{\mathrm{B}}$/MeV\\
\hline
$^{40}$Ca + $^{40}$Ca  &  62.531 & 1.174 &  0.652  & 54.925  & 9.735   & 3.757 \\
$^{40}$Ca + $^{48}$Ca  &  64.925 & 1.174 &  0.657  & 53.189  & 10.080  & 3.503 \\
$^{48}$Ca + $^{48}$Ca  &  64.104 & 1.175 &  0.662  & 51.751  & 10.379  & 3.249 \\
$^{58}$Ni + $^{58}$Ni  &  72.816 & 1.177 &  0.671  & 99.413  & 10.553  & 3.791 \\
$^{58}$Ni + $^{64}$Ni  &  73.873 & 1.177 &  0.673  & 97.713  & 10.754  & 3.661 \\
$^{64}$Ni + $^{64}$Ni  &  73.845 & 1.177 &  0.676  & 96.178  & 10.939  & 3.524 \\
\bottomrule
\end{tabular*}
\vspace{0mm}
\end{center}
\vspace{0mm}
\begin{multicols}{2}

Further, Figure~\ref{ReCa} shows the variation of RE with the reduced energy, the ratio of the projectile energy in the center-of-mass frame ($E_{\mathrm{c.m.}}$) to the Coulomb barrier energy ($V_{\mathrm{B}}$), for $^{40,48}$Ca + $^{40,48}$Ca~\cite{Jia16}.
For $^{40}$Ca + $^{48}$Ca, the RE deviates from unity with decreasing energy at sub-barrier energy region.
This isotopic dependence of RE is a strong sign for the PQNT effect in the sub-barrier fusion of $^{40}$Ca + $^{48}$Ca, considering the fact that the inelastic couplings of the reactants which should be considered in the CC calculations have been calibrated by using the experimental data of $^{40}$Ca + $^{40}$Ca and $^{48}$Ca + $^{48}$Ca.
A similar behavior also occurs for $^{58,64}$Ni + $^{58,64}$Ni in Fig.~\ref{ReNi}, although the data quality of the experimental cross sections should be improved for such an quantitative analysis.
This analysis gives a quantitative estimation for the effect of the sub-barrier fusion enhancement correlated with PQNT.

Also, it seems that the WS potential works well in the whole measured energy region, at least, for the fusion of the studied symmetrical systems of Ca + Ca and Ni + Ni.

Moreover, one matter should be pointed out from Figs.~\ref{ReCa} and~\ref{ReNi} is that the RE for both $^{40}$Ca + $^{48}$Ca and $^{58}$Ni + $^{64}$Ni shows a decrease tendency below the peak-energy with decreasing energy below the barriers.
This decrease possibly means the fading out of the inelastic and/or PQNT coupling effect with the decreasing reaction energy.
This tendency is consistent with the qualitative physical image that the reaction channels will gradually shut down with decreasing reaction energy~\cite{Lin03}.
This phenomenon is highlighted here by using this method and may offer further constraint for the relevant theories.
The deep-seated reaction dynamics should be excavated further both experimentally and theoretically.

\begin{center}
\includegraphics[width=3.2in]{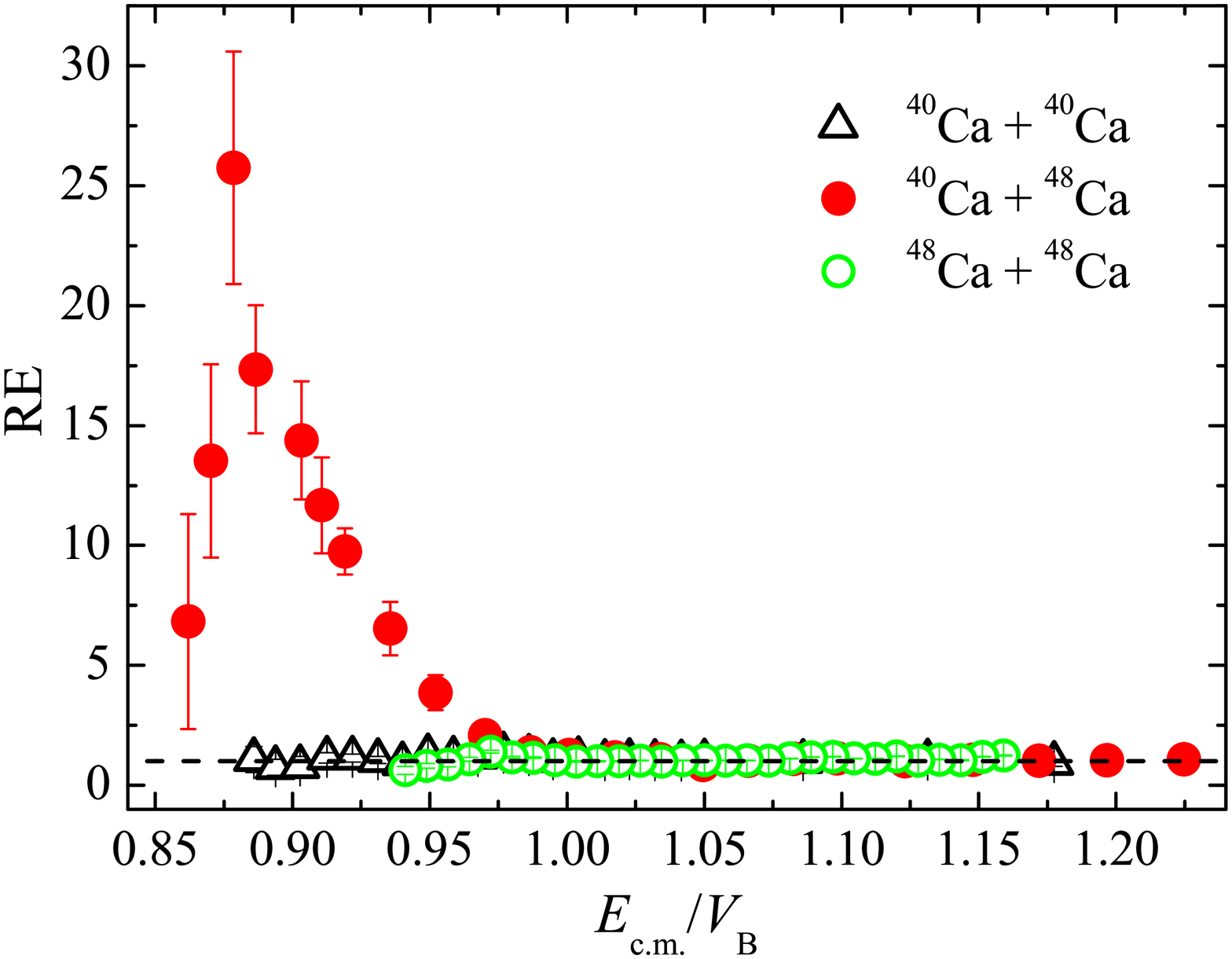}
\figcaption{\label{ReCa} (Color online) RE for $^{40,48}$Ca + $^{40,48}$Ca.
The original experimental fusion data are taken from Refs.~\cite{Montagnoli12,Trotta01,Jiang10}.}
\end{center}

\begin{center}
\includegraphics[width=3.2in]{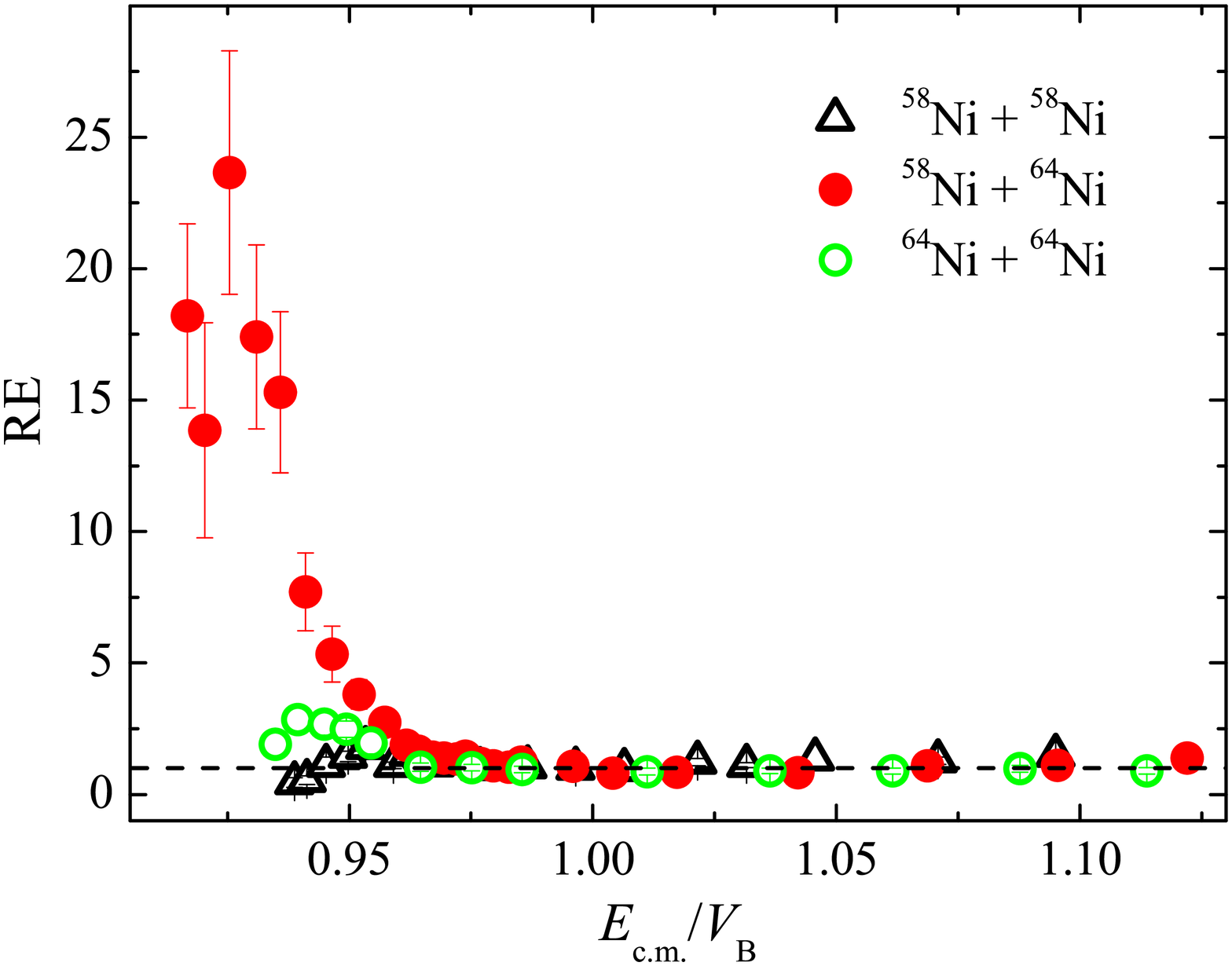}
\figcaption{\label{ReNi} (Color online) RE for $^{58,64}$Ni + $^{58,64}$Ni. The original experimental fusion data are taken from Refs.~\cite{Beckerman80}.}
\end{center}

\section{Summary}

In summary, in order to extract the PQNT effect quantitatively in near-barrier heavy-ion fusion, the self-consistent inelastic-constrained analysis with the RE method is further applied to analyze the typical fusion excitation functions of Ca + Ca and Ni + Ni.
The AW nuclear potential is used in the calculations.
This analysis shows that the CCFULL calculation result can be used as a scale for the RE, but still needs a reference for calibrating the inelastic coupling scheme in the CC calculations for the involved nuclei.

This analysis gives a further evidence for the sub-barrier fusion enhancement correlated with PQNT.
Additionally, the quantitative fusion enhancement extent is also extracted by using the RE method which should be helpful for checking and promoting the current theoretical models.

\end{multicols}

\vspace{-1mm}
\centerline{\rule{80mm}{0.1pt}}
\vspace{2mm}

\begin{multicols}{2}

\end{multicols}

\clearpage
%\end{CJK*}
\end{document}